\documentclass[aps,prb,reprint,superscriptaddress,amsmath,amssymb,twocolumn]{revtex4-2}
\usepackage{graphicx}% Include figure files
\usepackage{dcolumn}% Align table columns on decimal point
\usepackage{bm}% bold math
\usepackage{amsmath}
\usepackage{microtype}% Fine tune type setting
\usepackage{array}
\usepackage{float}
\usepackage{dcolumn,graphicx,color,booktabs,microtype,afterpage}
\usepackage[charter,greekuppercase=italicized]{mathdesign}
\usepackage{sidecap}
\usepackage[mathlines]{lineno}

\usepackage[usenames,dvipsnames]{xcolor}
\definecolor{nblue}{rgb}{0.0, 0.0, 1.0}
\definecolor{blue}{rgb}{0.0, 0.0, 1.0}
\definecolor{magenta}{rgb}{0.79, 0.08, 0.48}
\usepackage[colorlinks,plainpages=false,linkcolor=blue,urlcolor=blue,citecolor=blue,pdfpagemode=UseNone,pdfstartview=FitBH]{hyperref}
\definecolor{orange}{rgb}{1, 0.5, 0}

\newcommand{\beq}{\begin{equation}}
	\newcommand{\eeq}{\end{equation}}
\newcommand{\bea}{\begin{eqnarray}}
	\newcommand{\eea}{\end{eqnarray}}

\newcommand{\tcr}[1]{{\color[rgb]{0,0,0}{#1}}}
\newcommand{\tcb}[1]{{\color[rgb]{0,0,0.8}{#1}}}

\renewcommand{\figurename}{Fig.}
\renewcommand{\tablename}{Table}
\makeatletter\renewcommand{\fnum@figure}[1]{\textbf{\figurename~\thefigure\, }}\makeatother
\makeatletter\renewcommand{\fnum@table}[1]{\textbf{\tablename~\thetable\, }}\makeatother

\begin{document}

\makeatletter\renewcommand{\ps@plain}{%
\def\@evenhead{\hfill\itshape\rightmark}%
\def\@oddhead{\itshape\leftmark\hfill}%
\renewcommand{\@evenfoot}{\hfill\small{--~\thepage~--}\hfill}%
\renewcommand{\@oddfoot}{\hfill\small{--~\thepage~--}\hfill}%
}\makeatother\pagestyle{plain}

\preprint{\textit{Preprint: \today}} 

\title{Large unconventional anomalous Hall effect far above room temperature in\\ epitaxial Fe$_3$Ga$_4$ films}

\author{Jing Meng}
\affiliation{Key Laboratory of Polar Materials and Devices (MOE), School of Physics and Electronic Science, East China Normal University, Shanghai 200241, China} 
\author{Huali Yang}
\affiliation{Key Laboratory of Magnetic Materials Devices, Zhejiang Province Key Laboratory of Magnetic Materials and Application Technology, Ningbo Institute of Materials Technology and Engineering, Chinese Academy of Sciences, Ningbo 315201, China} 
\author{Yu Shen}
\affiliation{Key Laboratory of Polar Materials and Devices (MOE), School of Physics and Electronic Science, East China Normal University, Shanghai 200241, China} 
\author{Kun Zheng}
\affiliation{Key Laboratory of Polar Materials and Devices (MOE), School of Physics and Electronic Science, East China Normal University, Shanghai 200241, China} 
\author{Hongru Wang}
\affiliation{Key Laboratory of Polar Materials and Devices (MOE), School of Physics and Electronic Science, East China Normal University, Shanghai 200241, China} 
\author{Yuhao Wang}
\affiliation{Key Laboratory of Polar Materials and Devices (MOE), School of Physics and Electronic Science, East China Normal University, Shanghai 200241, China} 
\author{Keqi Xia}
\affiliation{Key Laboratory of Polar Materials and Devices (MOE), School of Physics and Electronic Science, East China Normal University, Shanghai 200241, China} 
\author{Bocheng Yu}
\affiliation{Key Laboratory of Polar Materials and Devices (MOE), School of Physics and Electronic Science, East China Normal University, Shanghai 200241, China}
\author{Xiaoyan Zhu}
\affiliation{Key Laboratory of Polar Materials and Devices (MOE), School of Physics and Electronic Science, East China Normal University, Shanghai 200241, China}
\author{Baiqing Lv}
\affiliation{Tsung-Dao Lee Institute, Zhangjiang Institute for Advanced Study, School of Physics and Astronomy, Shanghai Jiao Tong University, Shanghai 200240, China}
\author{Yaobo Huang}
\affiliation{Shanghai Synchrotron Radiation Facility, Shanghai Advanced Research Institute, Chinese Academy of Sciences, Shanghai 201204, China}
\author{Jie Ma}
\affiliation{Key Laboratory of Artificial Structures and Quantum Control, School of Physics and Astronomy, Shanghai Jiao Tong University, Shanghai 200240, China}
\author{Dariusz Jakub Gawryluk}
\affiliation{PSI Center for Neutron and Muon Sciences CNM, CH-5232 Villigen PSI, Switzerland}
\author{Toni Shiroka}
\affiliation{PSI Center for Neutron and Muon Sciences CNM, CH-5232 Villigen PSI, Switzerland}
\affiliation{Laboratorium f\"ur Festk\"orperphysik, ETH Z\"urich, CH-8093 Z\"urich, Switzerland}
\author{Zhenzhong Yang}\email[Corresponding author:\\]{zzyang@phy.ecnu.edu.cn}
\affiliation{Key Laboratory of Polar Materials and Devices (MOE), School of Physics and Electronic Science, East China Normal University, Shanghai 200241, China}
\author{Yang Xu}
\affiliation{Key Laboratory of Polar Materials and Devices (MOE), School of Physics and Electronic Science, East China Normal University, Shanghai 200241, China}
\author{Qingfeng Zhan}\email[Corresponding author:\\]{qfzhan@phy.ecnu.edu.cn}
\affiliation{Key Laboratory of Polar Materials and Devices (MOE), School of Physics and Electronic Science, East China Normal University, Shanghai 200241, China}
\author{Tian Shang}\email[Corresponding author:\\]{tshang@phy.ecnu.edu.cn}
\affiliation{Key Laboratory of Polar Materials and Devices (MOE), School of Physics and Electronic Science, East China Normal University, Shanghai 200241, China}
\affiliation{Chongqing Key Laboratory of Precision Optics, Chongqing Institute of East China Normal University, Chongqing 401120, China}

\begin{abstract}
%%% limited at 150 words only
\noindent\textbf{Abstract}\\ 
Noncoplanar spin textures usually exhibit a finite scalar spin chirality (SSC) that can generate effective magnetic fields and lead to additional contributions to the Hall effect, namely topological or unconventional anomalous Hall effect (UAHE).
Unlike topological spin textures (e.g., magnetic skyrmions), materials that exhibit fluctuation-driven SSC and UAHE are rare.   
So far, their realization has been
limited to either low temperatures or high magnetic fields, both of which are unfavorable for
practical applications. 
Identifying new materials that exhibit UAHE in a low magnetic
field at room temperature is therefore essential.
Here, we report the discovery of a large UAHE far above room temperature
in epitaxial Fe$_3$Ga$_4$ films, where the fluctuation-driven SSC 
stems from the field-induced transverse-conical-spiral phase. 
Considering their epitaxial nature and the large UAHE stabilized
at room temperature in a low magnetic field, Fe$_3$Ga$_4$ films
represent an exciting, albeit rare, case of a promising
candidate material for spintronic devices.
%% 148 words
\end{abstract}

\maketitle\enlargethispage{3pt}

\noindent\textbf{Introduction}\\
Topological transport is a powerful yet simple method to
probe the nontrivial spin textures in magnetic materials~\cite{Nagaosa2020,Wang2022,Liu2024}. As such, it has become a hot topic
of frontier research in condensed matter physics and materials science.
A typical example of topological transport is the topological
Hall effect (THE), which has been frequently observed in materials with
topological spin textures, such as magnetic skyrmions~\cite{Fert2013,Kanazawa2017,Reichhardt2022,Fert2017,Nagaosa2013,Jiang2017b,Tokura2021}.
The nontrivial topology of skyrmions implies a Berry curvature,
which acts as a virtual magnetic field and gives rise to an additional
transverse contribution to the Hall effect of conduction electrons, namely,
to the THE~\cite{Nagaosa2013,Jiang2017b,Tokura2021}. 
Due to its ease of measurement, THE can be a convenient tool for the 
electrical read-out of spin textures in real applications~\cite{Maccariello2018}.
Besides skyrmions, magnetic materials with noncoplanar or noncollinear spin
textures may also show an unconventional anomalous Hall effect (UAHE).
This appears as an addition to the conventional component that is
induced by the spontaneous magnetization through the spin–orbit interaction
~\cite{Taguchi2001,Machida2007,Ueda2012,Neill2019,Giri2020,Rout2019,Roychowdhury2024,Skorupskii2024,Nagaosa2010}, challenging the necessity of topological spin textures for the THE or UAHE.
Note that, in some previous work, the observed UAHE was also 
refered to as THE~\cite{Neill2019,Ueda2012,Giri2020,Rout2019,Roychowdhury2024}.  
To avoid confusion, UAHE will be used here to refer to the
magnetic materials \emph{without} topological spin textures.   
As an alternative to the real-space scenario, UAHE has also been
observed in systems with band structure anomalies,
such as Weyl points near the Fermi level, which carry a significant
Berry curvature in momentum space~\cite{Nakatsuji2015,Liang2015}.
Similar to the case of magnetic skyrmions, the scalar spin chirality (SSC) $\chi_\mathrm{ijk}$ = $\boldsymbol{S_\mathrm{i}}$ $\cdot$ ($\boldsymbol{S_\mathrm{j}}$ $\times$ $\boldsymbol{S_\mathrm{k}}$), which represents the solid angle subtended by adjacent spins and determines the virtual magnetic field~\cite{Nagaosa2013,Jiang2017b,Tokura2021}, can also be finite in some
noncoplanar magnetic materials and can be coupled to the conduction electrons to produce a UAHE.

Recent theoretical work has shown that a finite SSC can also arise 
dynamically from thermal spin fluctuations and persist at temperatures above the magnetic order. In such cases, the enhanced UAHE can be understood in terms of both skew scattering~\cite{Ishizuka2018,Yasuyuki2019,Yang2020} and Berry curvature~\cite{Hou2017,Lu2019,Kolincio2021}. 
Such fluctuation- and/or thermal-driven UAHE has been experimentally realized in kagome metals~\cite{Ghimire2020,Fruhling2024,Zhang2022,Kolincio2021,Kolincio2023}, cobaltates~\cite{Abe2024}, chiral magnets or ferromagnetic metals~\cite{Gong2021,Roychowdhury2024,Fujishiro2021}, as well as in multilayer films~\cite{Wang2019,Cheng2019}.        
Nevertheless, in most the above cases, the prerequisites for potential applications are not yet met. The UAHE appears either at low temperatures, or at high magnetic fields, or only in a narrow temperature range close to the magnetic order.
For example, Pt/Cr$_2$O$_3$ bilayers with a Cr$_2$O$_3$ thickness
less than 6\,nm exhibit UAHE at temperatures above the N\'{e}el order ($\sim$300\,K) of Cr$_2$O$_3$~\cite{Cheng2019}, but the unconvenional anomalous Hall resistivity $\rho_\mathrm{xy}^\mathrm{U}$ is extremely small, less than 1\,n$\mathrm{\Omega}$\,cm.   
In this regard, it is essential to be able to identify new materials
exhibiting a large $\rho_\mathrm{xy}^\mathrm{U}$ at room temperature
in zero- or a very small field.

\begin{figure*}[!thp]
	\centering
	\includegraphics[width=0.85\textwidth,angle=0]{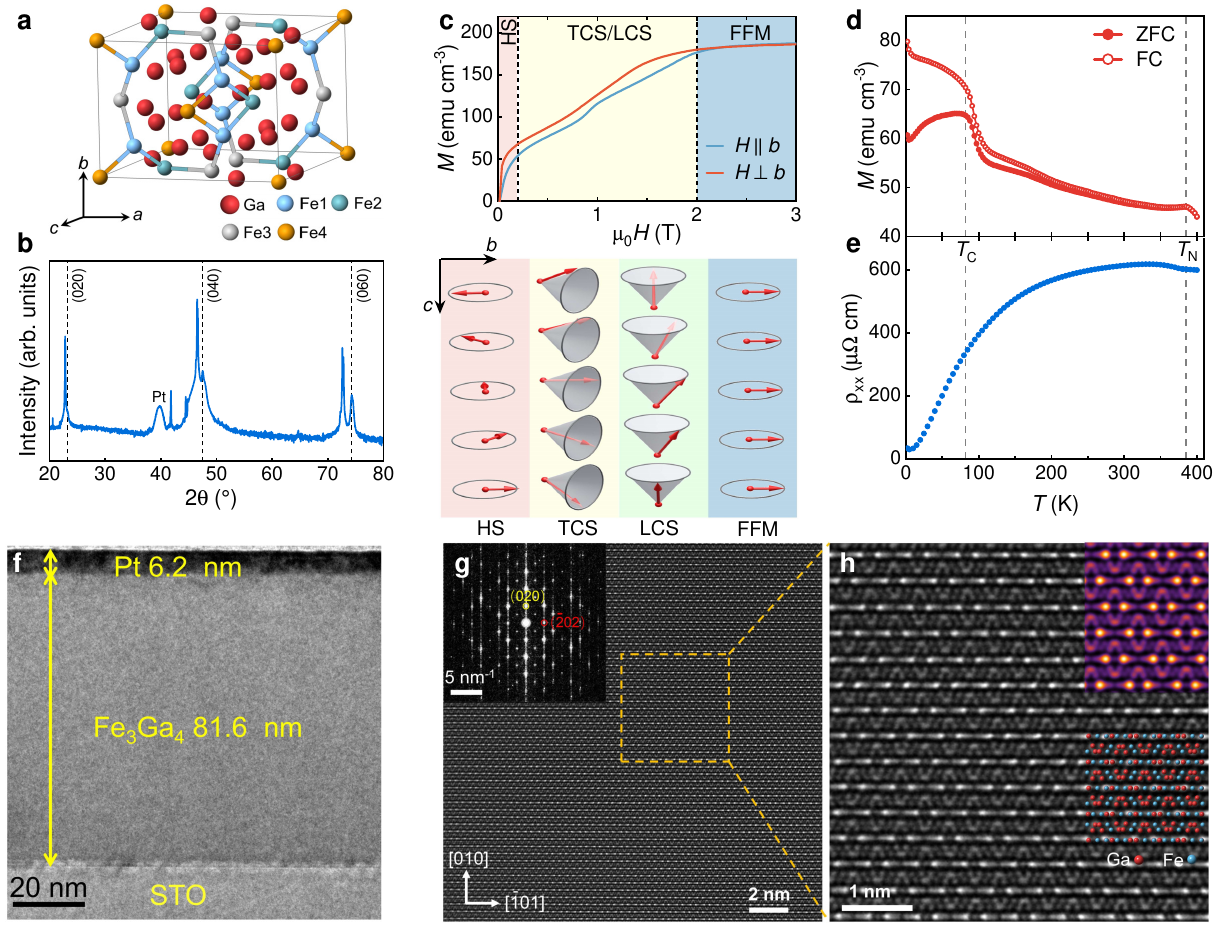}
	%\vspace{-2ex}%
	\caption{\label{fig:XRD}\textbf{Crystal structural and characterization of the epitaxial Fe$_3$Ga$_4$ film.} \textbf{a} Unit-cell crystal structure of monoclinic Fe$_3$Ga$_4$, highlighting the four nonequivalent Fe and Ga sites. 
	Here, only the Fe atoms are shown in different colors, while
	all the Ga atoms are shown in red.
		\textbf{b} Representative XRD pattern of a Pt/Fe$_3$Ga$_4$/SrTiO$_3$ film.
		The series of observed (0$k$0) reflections confirms that the Fe$_3$Ga$_4$ film was epitaxially grown along the $b$ axis on the (001)-oriented STO substrate. The sharp peaks at $2\theta \sim 21^\circ$ and 42$^\circ$ are attributed to the STO substrate (see Supplementary Fig.~S1). \textbf{c} In-plane and out-of-plane field-dependent magnetization of a Fe$_3$Ga$_4$ film collected at 200\,K. Various magnetic phases, including HS, TCS, LCS, and FFM order, appear as the magnetic field increases. Their magnetic  
		structures are schematically shown below the panel. In the HS phase, the magnetic order propagates along the $c$ axis, with the Fe moments rotating in the $ab$ plane. For $H \parallel b$ ($H \perp b$), a metamagnetic transition from HS to TCS (LCS) was proposed.
		\textbf{d} Temperature-dependent zero-field-cooling (ZFC)- and field-cooling (FC) magnetization $M(T)$. 
		\textbf{e} Temperature-dependent electrical resistivity $\rho_\mathrm{xx}(T)$.   
		The $M(T)$ was collected in a field of 0.1\,T applied perpendicular to the film plane, while $\rho_\mathrm{xx}(T)$ was collected in a zero-field condition. The dashed lines mark the AFM and FM transition at $T_\mathrm{N}$ $\sim$ 385\,K and $T_\mathrm{C}$ $\sim$ 82\,K, respectively.
		\textbf{f} TEM image of the Pt/Fe$_3$Ga$_4$/STO film cross-section. The Pt and Fe$_3$Ga$_4$ layers, as well as the STO substrate are clearly visible with different contrasts. 
		\textbf{g} Atomic-resolution HAADF-STEM image and the corresponding FFT image, demonstrating the single-crystalline nature of Fe$_3$Ga$_4$ film.
		\textbf{h} The enlarged view of the orange-marked region in panel (g), overlaid with a simulated HAADF image and crystal models. No domain structures could be identified along the $b$-axis of the film. Blue and red spheres represent Fe and Ga atoms, respectively.}
\end{figure*}

The intermetallic Fe$_3$Ga$_4$ compound was reported to exhibit
a clear UAHE in a wide temperature range below $\sim$400\,K~\cite{Mendez2015}.
Fe$_3$Ga$_4$ adopts a centrosymmetric monoclinic structure ($C2/m$), whose unit cell contains 18 Fe and 24 Ga atoms (see Fig.~\ref{fig:XRD}a)~\cite{Kawamiya1986}. There are four crystallographically inequivalent Fe sites, 
which allows for a variety of possible nearest- and next-nearest neighbor interactions. 
In weak magnetic fields, Fe$_3$Ga$_4$ 
undergoes three magnetic transitions: a paramagnetic (PM) to ferromagnetic (FM) transition at $T_1$ $\sim$ 420\,K, a FM to antiferromagnetic (AFM) transition at $T_2$ $\sim$ 360\,K, and, finally, an AFM to the ground-state FM transition at $T_3$ $\sim$ 68\,K~\cite{Mendez2015}. 
In the intermediate AFM state ($T_3 < T < T_2$), an incommensurate
spin-density-wave (ISDW) or a helical spiral (HS) order are
equally consistent with the neutron diffraction data, with
the latter being more compatible with the theoretical predictions
and magnetization data~\cite{Wu2018,Mahdi2021,Wilfong2022}.
Depending on the orientation, an applied magnetic field transforms
the HS order into a transverse conical spiral (TCS) or a
longitudinal conical spiral (LCS) order, before entering the forced
ferromagnetic (FFM) state (see Fig.~\ref{fig:XRD}c)~\cite{Wilfong2022}. 
Similar to the kagome metal $RE$Mn$_6$Sn$_6$ ($RE$ = Sc, Y, Er)~\cite{Ghimire2020,Fruhling2024,Zhang2022}, the TCS order consists of a noncoplanar spin configuration that produces a finite $\chi_\mathrm{ijk}$ via
dynamic fluctuations, which could
account for the observed UAHE in Fe$_3$Ga$_4$~\cite{Mendez2015}.
However, the appearance of topological spin textures in Fe$_3$Ga$_4$
cannot be fully excluded.
To date, the origin of the UAHE, as well as its temperature- and
field evolution, remain largely unexplored. 
These rich magnetic phases, involving both the temperature and
the magnetic field, originate from the 
complex magnetic interactions~\cite{Mahdi2021,Wilfong2022}.
As a consequence, the magnetic properties of Fe$_3$Ga$_4$ can be effectively tuned via external stimulus, such as, the magnetic field, doping, pressure, annealing, and dimensionality~\cite{Mendez2015,Wilfong2022,Wilfong2021,Kanani1998,Kawamiya1986b,Wilfong2022b,Moura2016}. 

Here, we report on the epitaxial growth of Fe$_3$Ga$_4$ films on a SrTiO$_3$ (STO) substrate and the systematic study of their magnetic- and transport properties. The unconventional anomalous Hall effect is shown to occur
in the intermediate AFM phase, in the magnetic-field region where a series of metamagnetic transitions take place, and to cover a wide temperature range ($\sim$100--380\,K). The occurrence of UAHE in Fe$_3$Ga$_4$ films is most likely attributed to the fluctuation-driven finite SSC in the noncoplanar spiral states.
Besides $RE$Mn$_6$Sn$_6$, Fe$_3$Ga$_4$ is, to the best our knowledge, the second case, where chirality fluctuations originating from the spiral phase play a key role in determining the topological transport properties.
Overall, we find that the Fe$_3$Ga$_4$ film is among the most promising candidate platforms for spintronic applications.

%==== figure =============================%
%=== end figure ==========================%
%

\vspace{4pt}
\noindent\textbf{Results}\\
\noindent\textbf{Crystal structure and characterizations of Fe$_3$Ga$_4$ films}\\
The epitaxial nature of the sputtered Fe$_3$Ga$_4$ film was checked by x-ray diffraction (XRD) and transmission electron microscopy (TEM) measurements. Figure~\ref{fig:XRD}b shows a representative 2$\theta$--$\omega$
scan of a Fe$_3$Ga$_4$ film sputtered on a (001)-oriented STO substrate.
The (0$k$0) reflections of Fe$_3$Ga$_4$ were identified near the (00$l$) reflections of the STO substrate, implying that the Fe$_3$Ga$_4$ was epitaxially grown along the $b$ axis.
No indication of impurities or misorientations was detected in the
XRD pattern. The epitaxial nature of Fe$_3$Ga$_4$ film was
further confirmed by %the
reciprocal space mapping (see Supplementary Fig.~S1).
The determined out-of-plane lattice parameter $b$ = 7.602~\AA{} of
the film is comparable to its bulk value~\cite{Mendez2015}. 
The cross-sectional TEM image reveals distinct interfaces between the
Pt capping layer, the Fe$_3$Ga$_4$ film, and the STO substrate
(see Fig.~\ref{fig:XRD}f), with estimated thicknesses of
6.2 and $\sim$81.6\,nm for the Pt layer and the Fe$_3$Ga$_4$ film,
respectively. 
To check the crystal structure at a microscopic level, a cross sectional high-angle annular dark field-scanning transmission electron microscopy (HAADF-STEM) measurement was also performed. Figure~\ref{fig:XRD}g illustrates the HAADF-STEM image of Fe$_3$Ga$_4$ film viewed along the [101] direction. Clear diffraction spots in the fast Fourier transform (FFT) image (shown in the inset) suggest a well-crystallized Fe$_3$Ga$_4$ film. Both Fe and Ga atoms can be well identified in the HAADF-STEM image (see the enlarged plot in Fig.~\ref{fig:XRD}h), consistent with the monoclinic crystal structure of Fe$_3$Ga$_4$.

The electronic 
properties of the Fe$_3$Ga$_4$ film were characterized by both magnetization
$M$ and electrical resistivity $\rho_\mathrm{xx}$ measurements.
The temperature-dependent magnetization $M(T)$ in Fig.~\ref{fig:XRD}d
shows two distinct transitions at $T_\mathrm{C} = 82$\,K and
$T_\mathrm{N} = 385$\,K, which correspond to the FM ($T_3$) and AFM ($T_2$)
transitions of the bulk crystal~\cite{Mendez2015,Wu2018,Wilfong2022}, respectively.
The sharp magnetic transition at $T_\mathrm{N}$ (with $\Delta T \sim 7$\,K) confirms
the high quality of the epitaxial Fe$_3$Ga$_4$ film, which is rare among the 
high-temperature helimagnetic films.
The AFM transition is reflected also in the $\rho_\mathrm{xx}(T)$ data
(see dashed lines in Fig.~\ref{fig:XRD}e), but the FM transition is
less evident (see Supplementary Fig.~S2). The residual resistivity
ratio (RRR $\sim$ 17.7) is much larger than that of % OR \tcr{in}
the bulk crystal~\cite{Mendez2015,Wu2018,Wilfong2022}, again indicating
the high quality of our Fe$_3$Ga$_4$ film. The $M(T)$ data were collected
under various magnetic fields (up to 3\,T), applied both perpendicular ($H \parallel b$) and parallel ($H \parallel ac$) to the film plane (see Supplementary Fig.~S3). For $H \parallel b$, as the magnetic field increases, $T_\mathrm{C}$ shifts to $\sim 110$\,K for $\mu_0H$ = 1\,T. In applied fields above 1\,T, the FM transition evolves into a broad crossover, 
where the Fe moments start to be polarized along the field direction. By contrast, for $\mu_0H$ = 3\,T, $T_\mathrm{N}$ is continuously suppressed to 362\,K. The determined $T_\mathrm{C}$ and $T_\mathrm{N}$ values are summarized in the magnetic phase diagram (see below).  
A similar behavior was also found for $H \parallel ac$ (see details in Supplementary Fig.~S3).
Akin to the case of bulk crystals, also in Fe$_3$Ga$_4$ films the
temperature range where the intermediate AFM phase occurs is
suppressed by 
an external field~\cite{Mendez2015,Wilfong2022}.
Since the highest temperature (i.e., 400\,K) accessible in this work is lower than $T_1$, most of the measurements were performed in the low-$T$ FM and intermediate AFM states.

%=== begin figure ==========================%
\begin{figure*}[!thp]
	\centering
	\includegraphics[width=0.9\textwidth,angle= 0]{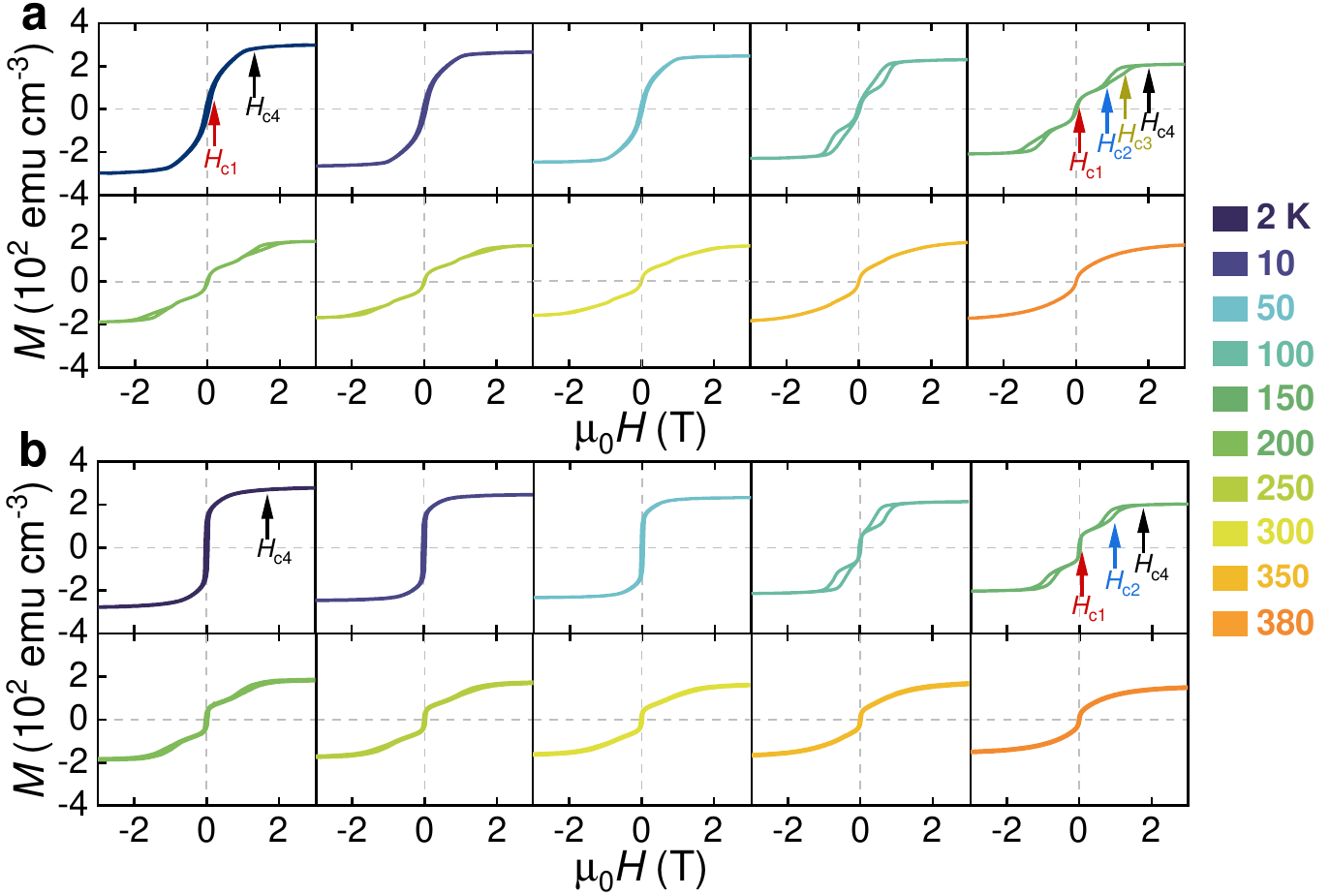}
	\caption{\label{fig:MH} \textbf{Magnetization of a Fe$_3$Ga$_4$ film}. \textbf{a,b}
		Field-dependent magnetization collected at various temperatures between 2 and 380\,K with the magnetic fields applied perpendicular (i.e., $H \parallel b$) (a) and parallel to the film plane  (i.e., $H \parallel ac$) (b). 
		The signal from the STO substrate was subtracted (see Supplementary Fig.~S4). The arrows mark the saturation field ($H_\mathrm{c4}$) and the three critical fields ($H_\mathrm{c1}$, $H_\mathrm{c2}$, and $H_\mathrm{c3}$), where the Fe$_3$Ga$_4$ film undergoes metamagnetic transitions. In the FM state, the metamagnetic transition was observed only for  $H \parallel b$. While, in the AFM state, we observe three metamagnetic transitions for $H \parallel b$, but only two transitions for $H \parallel ac$ (see the determination of $H_\mathrm{c1}$ to $H_\mathrm{c4}$ in Supplementary Figs.~S5 and S6).}
		%\tcr{It is noted that $H_\mathrm{c1}$ is most likely extrinsic and is attributed to the uncompensated magnetization at the interfaces (see  details in Supplementary Fig.~S8).}
\end{figure*}
%=== end figure ==========================%  
%

%=== begin figure ==========================%
\begin{figure*}[!thp]
	\centering
	\includegraphics[width=0.90\textwidth,angle= 0]{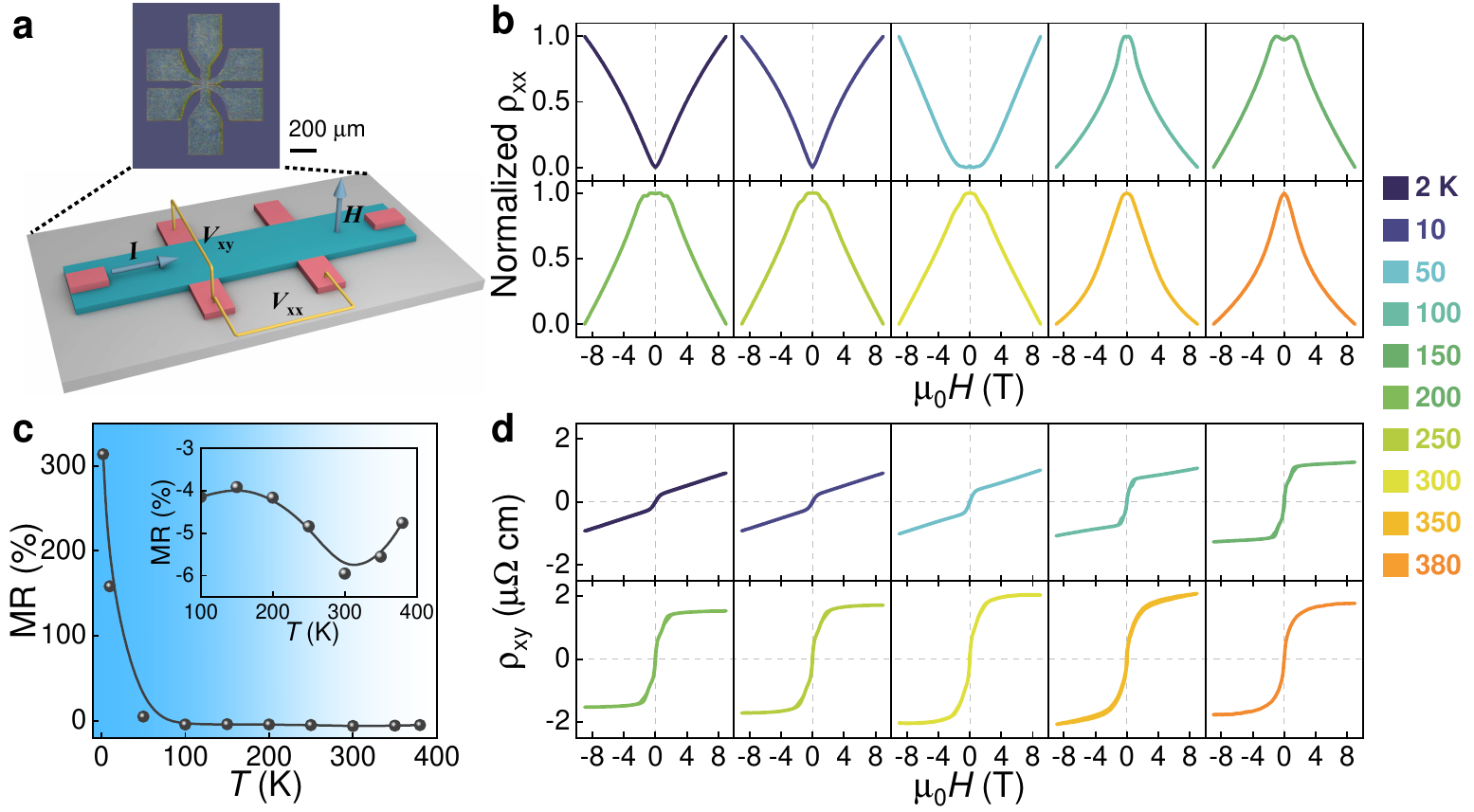}
	\caption{\label{fig:pH} \textbf{Transport properties of a Fe$_3$Ga$_4$ film}. 
		\textbf{a} A representative optical image of the micro-fabricated Pt/Fe$_3$Ga$_4$/STO film with a Hall-bar geometry. The schematic drawing below the panel depicts the Hall-bar device with six electrodes and notations of longitudinal electrical resistivity $\rho_\mathrm{xx}$ (= $V_\mathrm{xx}$/$I$) and transverse Hall resistivity $\rho_\mathrm{xy}$ (= $V_\mathrm{xy}$/$I$) measurements. For both measurements, the dc current was applied within the film plane ($I \parallel ac$), while the magnetic field was applied perpendicular to the film plane ($H \parallel b$).
		\textbf{b} Field-dependent electrical resistivity  $\rho_\mathrm{xx}(H)$ collected at various temperatures between 2 and 380\,K. 
		To better compare the $\rho_\mathrm{xx}(H)$ data at different temperatures, they were normalized to the 0--1 range.
		\textbf{c} Temperature dependence of the 9-T MR. The inset shows the enlarged view of MR in the AFM state. 
		\textbf{d} Field-dependent Hall resistivity $\rho_\mathrm{xy}(H)$ collected at various temperatures between 2 and 380\,K.
	}
\end{figure*}
%=== end figure ==========================%  

%
\vspace{5pt}
\noindent\textbf{Magnetization and metamagnetic transitions}\\
Previous work on bulk Fe$_3$Ga$_4$ crystals has shown that, 
in the intermediate AFM phase, the metamagnetic transitions are temperature-,
field-, and orientation dependent~\cite{Mendez2015,Wu2018,Wilfong2022}. 
Here we find that such metamagnetic transitions persist
also in Fe$_3$Ga$_4$ films.
After subtracting the substrate signal (see Supplementary Fig.~S4),
we obtain the field-dependent magnetization $M(H)$ at various temperatures,
as shown in Figs.~\ref{fig:MH}a,b for $H \parallel b$ and $H \parallel ac$, respectively. 
The magnetization saturates when the external field exceeds
the saturation field $H_\mathrm{c4}$, whose value 
varies across the different magnetic phases (see phase diagram below). 
For $H \parallel b$, in the AFM phase, the Fe$_3$Ga$_4$ film undergoes three metamagnetic transitions as the field increases. At each transition, $M(H)$ shows a small yet clear hysteresis, which becomes less evident when approaching the FM state. For $H < H_\mathrm{c4}$, as indicated by the arrows in the $M(H)$ {dataset at 150\,K, the three identified critical fields (determined by d$M$/d$H$ in Supplementary Figs.~S5 and S6) are $\mu_0H_\mathrm{c1}$ $\sim$ 0.04\,T, $\mu_0H_\mathrm{c2}$ $\sim$ 0.86\,T, and $\mu_0H_\mathrm{c3}$ $\sim$ 1.36\,T. 
While in the low-$T$ FM state, before the Fe moments are fully polarized at $H > H_\mathrm{c4}$, there is also a weak metamagnetic transition at $\mu_0H_\mathrm{c1}$ $\sim$ 0.08\,T (indicated by an arrow in the 2-K $M(H)$ dataset), which was also present in the bulk crystal~\cite{Wilfong2022}. According to the neutron diffraction data, in bulk crystals in the FM state, the Fe moments are polarized along the $c$ axis~\cite{Wu2018}. This leads to a much larger magnetization at low fields when $H \parallel c$ axis than for $H \parallel a$ or $b$ axis~\cite{Wu2018,Wilfong2022}. 
In a Fe$_3$Ga$_4$ film at low fields, the in-plane magnetization
is much larger than the out-of-plane magnetization (see Supplementary Fig.~S7),
indicating that the Fe moments are aligned within the film plane (here, the $ac$ plane). 
Further, although only two metamagnetic transitions could be identified for $H \parallel ac$ (see Fig.~\ref{fig:MH}b and Supplementary Fig.~S6), the critical fields are comparable to those obtained for $H \parallel b$.
Note that, the identified metamagnetic transition at $H_\mathrm{c1}$ is most likely extrinsic and attributed to the uncompensated magnetization at the interfaces (see Supplementary Fig.~S8). This assignment is also substantiated by recent neutron scattering studies on Fe$_3$Ga$_4$ bulk single crystals~\cite{Baral2025}.
For this reason, $H_\mathrm{c1}$ will not be discussed further below.

%
%==== figure =============================%
\begin{figure*}[th]
	\centering
	\includegraphics[width=0.90\textwidth,angle=0]{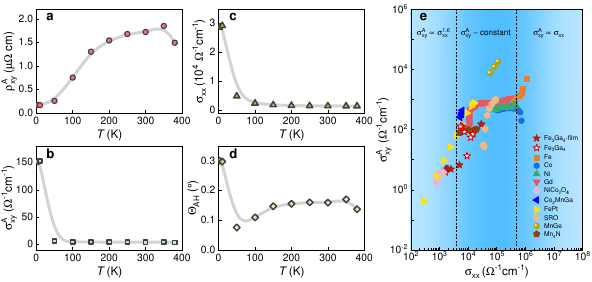}
	\vspace{-2ex}%
	\caption{\label{fig:AHE}% 
		\textbf{Anomalous Hall effect in a Fe$_3$Ga$_4$ film}. 
		\textbf{a}--\textbf{d} Temperature dependence of the anomalous Hall resistivity $\rho^\mathrm{A}_\mathrm{xy}(T)$ (a), the anomalous Hall conductivity $\sigma^\mathrm{A}_\mathrm{xy}(T)$ (b), the electrical conductivity $\sigma_\mathrm{xx}(T)$ (c), and the anomalous Hall angle $\Theta_\mathrm{A}(T)$ (d) for a Fe$_3$Ga$_4$ film. The analysis of Hall-resistivity data is reported in Supplementary Figs.~S10, S11, and S12.
		\textbf{e} $\sigma^\mathrm{A}_\mathrm{xy}$ vs %against the 
		$\sigma_\mathrm{xx}$ for various types of magnetic thin films spanning % the
		different AHE regimes, from the side-jump mechanism ($\sigma^\mathrm{A}_\mathrm{xy}$ $\propto$  $\sigma_\mathrm{xx}^{1.6}$), through the intrinsic- ($\sigma^\mathrm{A}_\mathrm{xy}$ $\sim$ const.), and skew-scattering  ($\sigma^\mathrm{A}_\mathrm{xy}$ $\propto$ $\sigma_\mathrm{xx}$) regimes. Except for the Fe$_3$Ga$_4$ film (solid stars), the data for other films were taken from Refs.~\cite{Miyasato2007,Kan2021,Park2020,Lu2013,Roy2023,Fujishiro2021,Isogami2021},
		while the data for the single-crystal bulk Fe$_3$Ga$_4$ (open
		stars) were taken from Ref.~\cite{Mendez2015}.}	
\end{figure*}
%=== end figure ==========================%
%

\vspace{5pt}
\noindent\textbf{Magnetoresistivity and Hall resistivity}\\
Metamagnetic transitions appear also in the field-dependent electrical resistivity $\rho_\mathrm{xx}(H)$. To perform the transport measurements, the Fe$_3$Ga$_4$ film was patterned into a Hall-bar geometry (see Fig.~\ref{fig:pH}a). 
In the FM state, there are no distinct anomalies in the $\rho_\mathrm{xx}(H)$ (Fig.~\ref{fig:pH}b), while both $H_\mathrm{c1}$ and $H_\mathrm{c4}$ can be identified in the d$\rho_\mathrm{xx}$/d$H$ data (see Supplementary Fig.~S9). In the AFM state, the metamagnetic transitions are obvious in the low-field region. All the critical fields, as determined from $\rho_\mathrm{xx}(H)$, are highly consistent with the $M(H)$ results (see phase diagram below).
Interestingly, in the low-$T$ FM state, Fe$_3$Ga$_4$ film exhibits a giant and positive magnetoresistivity (MR), reaching $\sim$ 313\% at 2\,K in a  field of 9\,T (see Fig.~\ref{fig:pH}c and Supplementary Fig.~S10). The MR decreases significantly with increasing temperature. Upon entering the intermediate AFM state, not only the sign of MR becomes negative, but also its magnitude is significantly reduced, dropping to  less than 10\%.  
A similar sign reversal of MR was observed also in bulk polycrystals~\cite{Duijn1998},
but it is absent in Fe$_3$Ga$_4$ single crystals~\cite{Mendez2015}. 
Such different types of MR behavior are most likely attributed to the magnetic anisotropy of Fe$_3$Ga$_4$. Further measurements in magnetic fields applied along different crystal orientations are desirable to clarify this issue.
Numerous metamagnetic transitions are frequently observed in magnetic
materials with nontrivial spin textures, such as
$RE$Mn$_6$Sn$_6$~\cite{Ghimire2020,Fruhling2024,Zhang2022}, SrCo$_6$O$_{11}$~\cite{Abe2024}, Gd$_3$Ru$_4$Al$_{12}$~\cite{Hirschberger2019}, MnSi~\cite{Thessieu1997}, and EuAl$_4$~\cite{takagi2022}, with the latter three known to host magnetic skyrmions.
In all these materials, the Hall resistivity is usually given by the sum of an ordinary ($\rho_\mathrm{xy}^\mathrm{O}$), an anomalous ($\rho_\mathrm{xy}^\mathrm{A}$), and  
an unconventional anomalous ($\rho_\mathrm{xy}^\mathrm{U}$) Hall contribution, namely, $\rho_\mathrm{xy} = \rho_\mathrm{xy}^\mathrm{O} + \rho_\mathrm{xy}^\mathrm{A} + \rho_\mathrm{xy}^\mathrm{U}$ 
~\cite{Taguchi2001,Machida2007,Ueda2012,Neill2019,Giri2020,Rout2019,Skorupskii2024,Roychowdhury2024,Nagaosa2013,Jiang2017b,Tokura2021}. 
Note that, in the literature, $\rho_\mathrm{xy}^\mathrm{U}$ is often (incorrectly) called topological Hall resistivity $\rho_\mathrm{xy}^\mathrm{T}$}.
Within the metamagnetic transitions, the above-mentioned materials exhibit
a $\rho_\mathrm{xy}^\mathrm{U}$ that results from the Berry phase acquired by the carriers when they pass through the nontrivial spin textures~\cite{Taguchi2001,Machida2007,Ueda2012,Neill2019,Giri2020,Rout2019,Skorupskii2024,Roychowdhury2024,Nagaosa2013,Jiang2017b,Tokura2021}. Also here we exploit the Hall-resistivity measurements on Fe$_3$Ga$_4$ films to search for possible nontrivial spin textures
(note, though, that this is only a necessary condition). In the low-$T$ FM state, $\rho_\mathrm{xy}(H)$ exhibits the typical features
due to 
the AHE in a ferromagnet (see Fig.~\ref{fig:pH}d). While in the intermediate AFM state,  $\rho_\mathrm{xy}(H)$ shows multiple kinks at the metamagnetic transitions at $H < H_\mathrm{c4}$ (see Supplementary Fig.~S11). As the magnetic field increases above 
$H_\mathrm{c4}$ (where the Fe moments are fully polarized), $\rho_\mathrm{xy}^\mathrm{A}$
becomes again dominant.

\vspace{5pt}
\noindent\textbf{Anomalous Hall effect}\\
The obtained $\rho_\mathrm{xy}^\mathrm{A}$ values versus temperature
are summarized in Fig.~\ref{fig:AHE}a, which reflects the different
magnetic states of a Fe$_3$Ga$_4$ film.
In the low-$T$ FM state, $\rho_\mathrm{xy}^\mathrm{A}$ is almost temperature-independent,
but it starts to increase when entering the intermediate AFM state.
As the temperature approaches the high-$T$ FM state, $\rho_\mathrm{xy}^\mathrm{A}$ starts to decrease again. Due to the reduced electrical resistivity (see Fig.~\ref{fig:XRD}e), both the electrical conductivity $\sigma_\mathrm{xx}$ and the anomalous Hall conductivity $\sigma_\mathrm{xy}^\mathrm{A}$ are greatly enhanced in the low-$T$ FM state (Fig.~\ref{fig:AHE}b,c). For instance, $\sigma_\mathrm{xy}^\mathrm{A} \sim 151$\,$\mathrm{\Omega}^{-1}$cm$^{-1}$ at 2\,K is almost 30 times larger than
the $\sigma_\mathrm{xy}^\mathrm{A}$ values at $T >$ 50\,K. The temperature-dependent anomalous Hall angle $\Theta_\mathrm{A}$ $\equiv$ $\tan$$^{-1}$($\sigma_\mathrm{xy}^\mathrm{A}/\sigma_\mathrm{xx}$) 
resembles $\sigma_\mathrm{xy}^\mathrm{A}(T)$, which exhibits clear anomalies close to $T_\mathrm{C}$ and $T_\mathrm{N}$ (Fig.~\ref{fig:AHE}d). $\Theta_\mathrm{A}$ is about 0.30$^\circ$ at 2\,K, a value 
comparable to that in the FM SrRuO$_3$ film or in the Fe$_3$Sn$_2$ kagome metal~\cite{Roy2023,Wang2016b}. In the intermediate AFM state, however, $\Theta_\mathrm{A}$ drops 
by a factor of 2. To elucidate the mechanism of AHE in the Fe$_3$Ga$_4$ film,
in Fig.~\ref{fig:AHE}e, we plot $\sigma_\mathrm{xy}^\mathrm{A}$ against
$\sigma_\mathrm{xx}$, together with the results from the Fe$_3$Ga$_4$
bulk crystal and of those of other magnetic thin films. 
The scaling relation between $\sigma_\mathrm{xy}^\mathrm{A}$ and $\sigma_\mathrm{xx}$ has been frequently studied in recent years. Generally, it can
be divided into three regimes, where different mechanisms have been
proposed to account for the $\sigma_\mathrm{xy}^\mathrm{A}$ behavior~\cite{Nagaosa2010,Chen2021,Yang2020}.
In the high-conductivity regime ($\sigma_\mathrm{xx}$ $\gtrsim$ 5 $\times$ 10$^5$\,$\mathrm{\Omega}^{-1}$cm$^{-1}$), the extrinsic skew scattering contribution dominates AHE and $\sigma_\mathrm{xy}^\mathrm{A}$ $\propto$ $\sigma_\mathrm{xx}$; in the good-metal regime (3 $\times$ 10$^3$ $\lesssim$ $\sigma_\mathrm{xx}$  $\lesssim$ 5 $\times$ 10$^5$\,$\mathrm{\Omega}^{-1}$cm$^{-1}$), $\sigma_\mathrm{xy}^\mathrm{A}$ is dominated by the intrinsic Berry-phase contribution, which is approximately independent of $\sigma_\mathrm{xx}$; finally in the bad-metal   
(or localized hopping) regime ($\sigma_\mathrm{xx}$ $\lesssim$ 3 $\times$ 10$^3$\,$\mathrm{\Omega}^{-1}$cm$^{-1}$), the extrinsic side-jump mechanism is at play and it leads to $\sigma_\mathrm{xy}^\mathrm{A}$ $\propto$  $\sigma_\mathrm{xx}^{1.6-1.8}$. In the Fe$_3$Ga$_4$ film, for $T < 400$\,K, we find that $\sigma_\mathrm{xx}$ $\sim$ 0.16--2.94 $\times$ 10$^4$\,$\mathrm{\Omega}^{-1}$cm$^{-1}$ (Fig.~\ref{fig:AHE}c) in both the good- and bad-metal regimes. Therefore, both the intrinsic Berry-phase- and the extrinsic side-jump mechanisms might account for the observed $\sigma_\mathrm{xy}^\mathrm{A}$ in the Fe$_3$Ga$_4$ film. Interestingly, $\sigma_\mathrm{xy}^\mathrm{A}$ in the low-$T$ FM state is compatible with 
the intrinsic regime. While, in the intermediate AFM state,
$\sigma_\mathrm{xy}^\mathrm{A}$ is mostly 
in the extrinsic regime. 
Different from the film, $\sigma_\mathrm{xy}^\mathrm{A}$ of a
Fe$_3$Ga$_4$ bulk crystal is located in the intrinsic regime
irrespective of its 
magnetic state. Such discrepancy is most likely attributed to the
additional disorder/scattering at the interfaces of a Fe$_3$Ga$_4$
film, resulting in an enhanced electrical resistivity.
It could be interesting to investigate Fe$_3$Ga$_4$ films with different
thicknesses, where the magnetic transition temperatures and the nature of $\sigma_\mathrm{xy}^\mathrm{A}$ can be tuned.

%
%==== figure =============================%
\begin{figure*}[th]
	\centering
	\includegraphics[width=0.95\textwidth,angle=0]{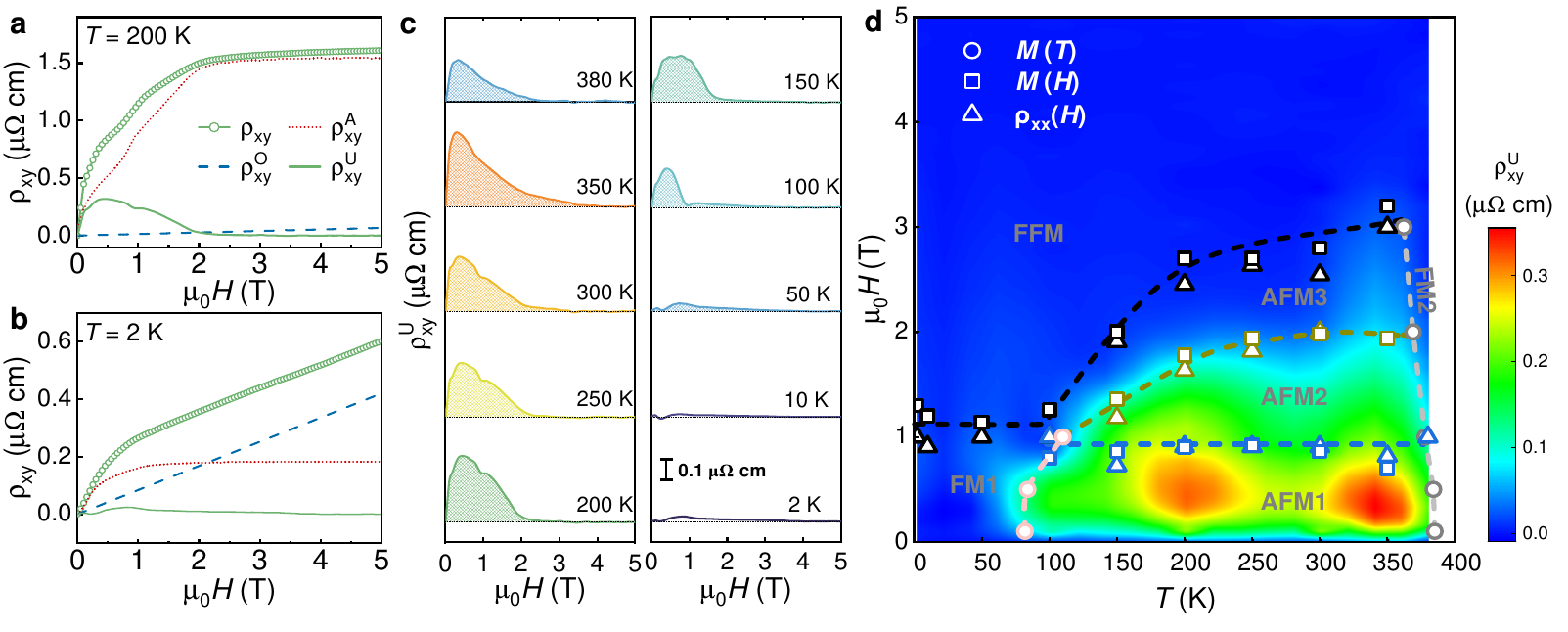}
	\vspace{-2ex}%
	\caption{\label{fig:THE}% 
	\textbf{%Topological 
	 Unconventional Hall effect and magnetic phase diagram of a Fe$_3$Ga$_4$ film}. \textbf{a-b} Analysis of the field-dependent Hall resistivity $\rho_\mathrm{xy}(H)$ collected at 200\,K (a) and 2 \,K (b) for a Fe$_3$Ga$_4$ film. Dashed, dotted, and solid lines represent the ordinary ($\rho_\mathrm{xy}^\mathrm{O}$), anomalous ($\rho_\mathrm{xy}^\mathrm{A}$), and unconventional anomalous ($\rho_\mathrm{xy}^\mathrm{U}$) contributions to the Hall resistivity, respectively.  
	The $\rho_\mathrm{xy}(H)$ curves at other temperatures were analyzed using the same procedure, (see Supplementary Fig.~S11).
	\textbf{c} Extracted field-dependent unconventional anomalous Hall resistivity $\rho_\mathrm{xy}^\mathrm{U}(H)$ at different temperatures between 2 and 380\,K. Here, only the data collected by increasing the magnetic field are plotted. 
	No visible 
	$\rho_\mathrm{xy}^\mathrm{U}$ could be identified in the low-$T$ FM state. 
	\textbf{d} Magnetic phase diagram of a Fe$_3$Ga$_4$ film with $H \parallel b$ (i.e., $H$ perpendicular to the film plane), as obtained from the $M(T)$, $M(H)$, and $\rho_\mathrm{xx}(H)$ datasets (see Supplementary Figs.~S3, S5, and S9). The background color represents the magnitude of $\rho_\mathrm{xy}^\mathrm{U}$ at various temperatures. AFM1, AFM2, and AFM3 denote the three different antiferromagnetic phases, while FM1 and FM2 are the two ferromagnetic phases. FFM stands for the forced ferromagnetic phase. Dashed lines in panel (d) are guides to the eyes.
	}	
\end{figure*}
%=== end figure =========================%

\vspace{5pt}
\noindent\textbf{%Topological 
Unconventional anomalous Hall effect far above room temperature}\\
To search for a possible unconventional anomalous contribution $\rho_\mathrm{xy}^\mathrm{U}$,
both $\rho_\mathrm{xy}^\mathrm{O}$ and $\rho_\mathrm{xy}^\mathrm{A}$ were %have to be
subtracted from the measured Hall resistivity $\rho_\mathrm{xy}$. 
While $\rho_\mathrm{xy}^\mathrm{O}$ (= $R_\mathrm{0}$$H$) is proportional
to the applied magnetic field, $\rho_\mathrm{xy}^\mathrm{A}$ (= $R_\mathrm{S}$$M$)
is mostly determined by the magnetization (see details in the Materials and Methods Section).
In real materials, $\rho_\mathrm{xy}^\mathrm{A}$ depends on the mechanism
at play, i.e., whether it is intrinsic-, side-jump, skew scattering, or
an intricate combination thereof~\cite{Nagaosa2010,Chen2021,Yang2020}.  
The different contributions to the Hall resistivity in the intermediate AFM state
(e.g., at 200\,K) and FM state of a Fe$_3$Ga$_4$ film are shown in
Fig.~\ref{fig:THE}a and Fig.~\ref{fig:THE}b, respectively, with the other temperatures showing similar results
(see Supplementary Fig.~S11).
Here, we use $\rho_\mathrm{xx}^2$ to extract
$\rho_\mathrm{xy}^\mathrm{A}$ by using the relation $\rho_\mathrm{xy}^\mathrm{A}$ = $S_\mathrm{H}\rho_\mathrm{xx}^2$$M$ in the AFM state, where $S_\mathrm{H}$ is a field-independent scaling factor. The obtained $R_0$ and $R_\mathrm{S}$ 
are summarized in Supplementary Fig.~S12.  
We note that the use of $S_\mathrm{H}\rho_\mathrm{xx}^2$, $S_\mathrm{H}$$\rho_\mathrm{xx}$, or simply $S_\mathrm{H}$ for $R_\mathrm{S}$ leads to a comparable unconventional anomalous contribution $\rho_\mathrm{xy}^\mathrm{U}(H)$ (see Supplementary Fig.~S13). 
After subtracting $\rho_\mathrm{xy}^\mathrm{O}(H)$ and $\rho_\mathrm{xy}^\mathrm{A}(H)$, the extracted unconventional anomalous contributions $\rho_\mathrm{xy}^\mathrm{U}(H)$ at various temperatures are shown in 
Fig.~\ref{fig:THE}c, as well as presented as a background in the magnetic phase diagram in Fig.~\ref{fig:THE}d.  
The hump-like anomaly in $\rho_\mathrm{xy}^\mathrm{U}(H)$ resembles the typical feature of UAHE caused by nontrivial spin textures~\cite{Taguchi2001,Machida2007,Ueda2012,Neill2019,Giri2020,Rout2019,Skorupskii2024,Roychowdhury2024,Nagaosa2013,Jiang2017b,Tokura2021}. $\rho_\mathrm{xy}^\mathrm{U}$ is almost absent in the low-$T$ FM state.
While in the intermediate AFM state, in the field range between $H_\mathrm{c1}$ and $H_\mathrm{c3}$ (marked as AFM1 and AFM2), 
where Fe$_3$Ga$_4$ undergoes metamagnetic transitions, $\rho_\mathrm{xy}^\mathrm{U}$ is particularly evident, and it becomes almost invisible at $H > H_\mathrm{c3}$ (i.e., AFM3 and FFM phases). 
The field-dependent $\rho_\mathrm{xy}^\mathrm{U}$ increases sharply and peaks at 
$\mu_0H$ $\sim$ 0.4\,T below the room temperature (Fig.~\ref{fig:THE}c). Such a peak moves to slightly lower fields when the temperature increases. $\rho_\mathrm{xy}^\mathrm{U}$ reaches a maximum value of $\sim$0.36\,{\textmu}$\mathrm{\Omega}$cm at 350\,K in a field of $\mu_0H$  = 0.35\,T. 
We found that the Fe$_3$Ga$_4$ film exhibits a much larger $\rho_\mathrm{xy}^\mathrm{U}$ than other  multilayer systems, which host topological spin textures at room temperature~\cite{Soumyanarayanan2017,He2018,Mourkas2021,Raju2019,Shaktiranjan2024}.  

\vspace{5pt}
\noindent\textbf{Discussion}\\
After comprehensively characterizing the magnetization and electrical transport of the Fe$_3$Ga$_4$ films, its magnetic phase diagram could be established (Fig.~\ref{fig:THE}d and Supplementary Fig.~S\tcb{16}). Similar to the bulk crystal~\cite{Mendez2015,Wilfong2022}, the Fe$_3$Ga$_4$ film also exhibits an intermediate AFM phase intercalated by a low-$T$ (FM1) and a high-$T$ FM phase (FM2).  
For $H \parallel b$, the Fe$_3$Ga$_4$ film undergoes two metamagnetic transitions before being fully polarized in the AFM phase at higher fields. While, for $H \parallel ac$, only  one metamagnetic transition can be identified. As the magnetic field increases, $\rho_\mathrm{xy}^\mathrm{U}$ increases sharply and peaks at $\mu_0H \sim$ 0.5\,T. Further, as the field is swept across $H_\mathrm{c2}$, $\rho_\mathrm{xy}^\mathrm{U}$ starts to decrease and it vanishes at $H > H_\mathrm{c3}$. Clearly, the UAHE appears only in the AFM1 and AFM2 phases (i.e., $H < H_\mathrm{c3}$) and across a very wide temperature range, i.e., $\sim 100$--380\,K.
The above features resemble those typical of materials with nontrivial spin textures~\cite{Skorupskii2024,Ghimire2020,Fruhling2024,Zhang2022,Abe2024,Hirschberger2019,Thessieu1997,takagi2022}, where UAHE or THE is attributed to the SSC, albeit mostly limited to much lower temperatures. Bulk Fe$_3$Ga$_4$ crystal possesses an HS order with a complex competition between FM and AFM interactions~\cite{Wu2018,Mahdi2021,Wilfong2022}, which also generates multiple metamagnetic transitions that are tuned by both magnetic field and temperature. The HS order evolves into an LCS- or a TCS noncoplanar order when applying a magnetic field parallel or perpendicular to the $c$ axis, respectively~\cite{Mahdi2021,Wilfong2022}.
In the case of Fe$_3$Ga$_4$ films, for $H \parallel b$ (see magnetic phase diagram in Fig.~\ref{fig:THE}d), the UAHE appears only in the AFM1 and AFM2 phases but is absent in the AFM3 phase. Thus, the former two phases are most likely to host a TCS state, while the latter is more likely to consist of a fan-like magnetic state, similar to that in $RE$Mn$_6$Sn$_6$ ($RE$ = Sc, Y) kagome metals~\cite{Ghimire2020,Zhang2022}.   

Now we discuss the possible mechanisms for the observed robust UAHE in Fe$_3$Ga$_4$ films. The THE has been found in a centrosymmetric breathing kagome lattice Gd$_3$Ru$_4$Al$_{12}$~\cite{Hirschberger2019} where, in the field range $\sim 1$--2\,T, a skyrmion lattice forms over the TCS phase below $\sim 10$\,K favored by thermal fluctuations. In the case of Fe$_3$Ga$_4$ films, the topological spin textures might also develop from its noncoplanar magnetic order. Fe$_3$Ga$_4$ films show a strong magnetic anisotropy (see Supplementary Fig.~S7), which can compete with its complex magnetic interactions and result in topological spin textures. Such a mechanism has been proposed to explain the formation of magnetic skyrmions in centrosymmetric GdRu$_2$Si$_2$~\cite{Khanh2020}. Similar to the Gd$_3$Ru$_4$Al$_{12}$ and GdRu$_2$Si$_2$, the Dzyaloshinskii-Moriya interaction (DMI) is also absent in the centrosymmetric Fe$_3$Ga$_4$ bulk crystal. However, DMI could be present in epitaxial Fe$_3$Ga$_4$ films.
Considering the rather thick Fe$_3$Ga$_4$ films ($\sim$81.6\,nm) investigated here, the effects of interfacial DMI at the Pt/Fe$_3$Ga$_4$
and Fe$_3$Ga$_4$/STO interfaces should be excluded. In general, to understand the role of interfacial DMI and its effects
on the topological transport properties, future studies of Fe$_3$Ga$_4$ films with different thickness (ideally less than 20\,nm) and different capping layers are highly desirable. However, the in-plane (i.e., $ac$-plane) lattice and angle mismatches between Fe$_3$Ga$_4$ film and STO substrate can cause inhomogeneous lattice distortions and domain structures (with a typical size of $\sim$200\,nm) in the film plane, both of which can locally break the inversion symmetry and thus, introduce a local DMI.

To gain more insight into the nature of observed UAHE, as well as to search for possible topological spin textures in Fe$_3$Ga$_4$ films,  preliminary magnetic force microscopy (MFM) measurements were carried out at room temperature. MFM is a commonly used technique for imaging nontrivial spin textures in real space as a complementary measurement to the topological transport properties~\cite{Soumyanarayanan2017,Raju2019}. 
The MFM measurements reveal distinct bubble-like magnetic domains, whose density increases progressively with the magnetic field (see Supplementary Fig.~S\tcb{17}). Interestingly, at room temperature, the estimated density $n_\mathrm{mb}(H)$
of magnetic bubbles scales almost linearly with $\rho_\mathrm{xy}^\mathrm{U}(H)$, which provides possible evidence that the appearance of topological spin textures could be at the origin of UAHE in the Fe$_3$Ga$_4$ films.
In the adiabatic approximation, that is, by assuming a strong coupling of the charge with a local spin, the topological Hall resistivity $\rho_\mathrm{xy}^\mathrm{T}$ is generally expressed as $\rho_\mathrm{xy}^\mathrm{T}$ = $P$$R_0$$B_\mathrm{eff}$.
Here, $P$ is the spin polarization of charge carriers, $R_0$ is the effective density of charge contributing to the THE, usually taken as the ordinary Hall coefficient, and $B_\mathrm{eff}$ is the effective (emergent) magnetic field experienced by carriers due to the Berry phase they accumulate while traversing the topological spin textures, e.g., skyrmions~\cite{Neubauer2009,Raju2019,Nagaosa2013}. The $B_\mathrm{eff}$ is defined as $B_\mathrm{eff}$ =  $n_\mathrm{sk}$$\Phi_0$, where $n_\mathrm{sk}$ is the skyrmion density and $\Phi_0$ the magnetic flux quantum. 
By assuming a spin polarization $P = 0.1$--1 and using the $n_\mathrm{mq}$ estimated from the MFM images, the calculated 
$\rho_\mathrm{xy}^\mathrm{T}$ is about 0.1--1$\times$10$^{-4}$\,{\textmu}$\mathrm{\Omega}$\,cm, which is thousand times smaller than the observed $\rho_\mathrm{xy}^\mathrm{U}$ = 0.26\,{\textmu}$\mathrm{\Omega}$\,cm at 300\,K. Such a significant discrepancy unambiguously excludes a key role of topological spin textures in determining the observed UAHE in Fe$_3$Ga$_4$ films.  
If any $\rho_\mathrm{xy}^\mathrm{T}$ exists, it should represent only a minor contribution to the observed total $\rho_\mathrm{xy}^\mathrm{U}$.  
However, the appearance of topological spin textures cannot be fully excluded, considering
the linear scaling of $\rho_\mathrm{xy}^\mathrm{U}(H)$ with $n_\mathrm{mb}(H)$ previously observed also in other skyrmion-hosting multilayers~\cite{Zeissler2018,Raju2019} and oxide thin films~\cite{Wang2018b,Vistoli2019}.
Thus, it could be interesting to further investigate the possible nontrivial spin textures in Fe$_3$Ga$_4$ films via other real-space imaging techniques~\cite{Zeissler2018,Romming2013,Pollard2017}.

As an alternative, the fluctuation-driven SSC is most likely the mechanism for UAHE in Fe$_3$Ga$_4$ films. Such a mechanism has been frequently used to explain the UAHE in different type of materials~\cite{Ghimire2020,Fruhling2024,Zhang2022,Kolincio2021,Kolincio2023,Abe2024,Gong2021,Roychowdhury2024,Fujishiro2021,Wang2019,Cheng2019}. Among these, the $RE$Mn$_6$Sn$_6$ family shows a similar picture to Fe$_3$Ga$_4$ films, where the 3$d$ magnetic moments of Mn or Fe undergo transitions from a helical spiral to a transverse-conical spiral and, finally, to
a fan-like state before entering the forced ferromagnetic state~\cite{Ghimire2020,Fruhling2024,Zhang2022,Wilfong2022}. In both $RE$Mn$_6$Sn$_6$ and Fe$_3$Ga$_4$, the UAHE appears solely in the TCS phase, which belongs to one of the noncoplanar spin structures that can produce a finite SSC with the assistance of dynamic fluctuations or in presence of DMI~\cite{Ghimire2020}.
Indeed, recently, such a fluctuation-driven mechanism has been
verified in Fe$_3$Ga$_4$ bulk crystals and shown to quantitatively
reproduce the observed UAHE~\cite{Mendez2015,Mahdi2021}.
Considering that both epitaxial Fe$_3$Ga$_4$ films and bulk crystal
show comparable magnetic phase diagrams and UAHE, the fluctuation-driven mechanism should be at the origin of the observed UAHE in the film as well. 
Our work also demonstrates that, in view of the relatively large electrical conductivity
(see Fig.~\ref{fig:AHE}e), the large UAHE of the Fe$_3$Ga$_4$
films is mostly related to the Berry curvature rather than to
the skew scattering mechanism.
Finally, we compare the Fe$_3$Ga$_4$ films with other materials that exhibit a fluctuation-driven UAHE (see Supplementary Table~S1). For instance, in the $RE$Mn$_6$Sn$_6$ family~\cite{Ghimire2020,Fruhling2024,Zhang2022}, the UAHE appears also in a wide temperature range ($\sim$100--300\,K) with a maximum $\rho_\mathrm{xy}^\mathrm{U}$ reaching $\sim 1$--2\,{\textmu}$\mathrm{\Omega}$\,cm. However, the critical field required to reach the maximum $\rho_\mathrm{xy}^\mathrm{U}$ is relatively large
(about 1\,T). The LaMn$_2$Ge$_2$ ferromagnet also shows results comparable to those of $RE$Mn$_6$Sn$_6$~\cite{Roychowdhury2024}. In the case of thin films, although the critical field is relatively small,
the UAHE appears either at low temperatures (e.g., at $\sim$130\,K in SrRuO$_3$)~\cite{Wang2019}, or in a narrow temperature range close to the magnetic order (e.g., in Pt/Cr$_2$O$_3$ films)~\cite{Cheng2019}. In addition, the $\rho_\mathrm{xy}^\mathrm{U}$ is extremely small in these films, much less than 0.1\,{\textmu}$\mathrm{\Omega}$\,cm.
The epitaxial Fe$_3$Ga$_4$ film is clearly superior to the above
mentioned materials. Its UAHE is present over a wide temperature
range ($\sim 100$--380\,K), with a $\rho_\mathrm{xy}^\mathrm{U}$
up to 0.4\,{\textmu}$\mathrm{\Omega}$\,cm achieved under a magnetic field of less than 0.4\,T.
\tcr{Note that the uncompensated magnetization at the interfaces
has little influence on the extracted temperature- and field-dependent
$\rho_\mathrm{xy}^\mathrm{U}$ in the Fe$_3$Ga$_4$ film. Although the
magnitude of $\rho_\mathrm{xy}^\mathrm{U}$ changes slightly when
considering the interface magnetization, the main features and the
extracted parameters are quite robust (see Supplementary Figs.~S14 and S15).}	
	
%	By utilizing the bulk magnetization, the resulting field-dependent unconventional anomalous Hall resistivity xyU(H) curve is nearly identical to that derived from the total magnetization, albeit with a slightly reduced absolute value [see Fig.R2(b)]. It is important to note that isolating the exact interface contribution to the magnetization from the total measured magnetization is quite challenging 
%	intensity of $\rho_\mathrm{xy}^\mathrm{U}$ gradually increases as the sign of the coefficient of interface magnetization changes from positive to negative. The corresponding $R_0$ and $R_s$ are extracted (Supplementary Fig. S15). However, the obtained $\rho_\mathrm{xy}^\mathrm{U}(H,T)$ is almost identical by using different methods, which indicated  $\rho_\mathrm{xy}^\mathrm{U}$ is not qualitatively affected by the spontaneous magnetization on the interface.(see Supplementary Fig.~S14)}

\vspace{5pt}
%\noindent\textbf{Conclusion}\\
In summary, by systematic measurements of electrical transport and
magnetization, we could establish the magnetic phase diagram of epitaxial
Fe$_3$Ga$_4$ films. Similar to the bulk crystal, Fe$_3$Ga$_4$ films exhibit
a rich magnetic phase diagram, including FM and AFM orders, as well as multiple metamagnetic transitions. The UAHE appears in the intermediate AFM state,  covering a very wide temperature range, from 100 to 380\,K. Such an AFM state is consistent with the TCS phase observed in the Fe$_3$Ga$_4$ bulk crystal. Therefore, we believe that the fluctuation-driven finite SSC in the TCS phase promotes
a large UAHE in Fe$_3$Ga$_4$ films.
Further advantages include the
possibility of epitaxial growth of a Fe$_3$Ga$_4$ film on single-crystalline
substrates or the high tunability of its topological properties
through strain engineering or piezoelectric effect. 
Finally, the large UAHE achieved well above room temperature
and in small magnetic fields makes the Fe$_3$Ga$_4$ film a very
promising platform for spintronic applications.

\emph{Note added}. After the present manuscript was submitted, a
related work by Baral et al.~\cite{Baral2025} appeared, in which bulk
Fe$_3$Ga$_4$ crystals were studied via neutron scattering under
various magnetic fields. In this study, a nontrivial spin
texture was also proposed to occur in the magnetic phase where
$\rho_\mathrm{xy}^\mathrm{U}$ reaches its maximum value.

\vspace{10pt}
\noindent\textbf{Methods}\\
\noindent\textbf{Thin-film growth}\\ 
The lattice parameters of the bulk Fe$_3$Ga$_4$ crystal are $a$ = 10.0979~\AA{}, $b$ = 7.6670~\AA{}, and $c$ = 7.8733~\AA{}~\cite{Mendez2015}. Therefore, an SrTiO$_3$ substrate
matches very well the Fe$_3$Ga$_4$ lattice, since its lattice parameter ($a$ = 3.860~\AA{}) is almost identical to half of the Fe$_3$Ga$_4$ $b$ axis. Fe$_3$Ga$_4$ films were epitaxially grown on (001)-oriented STO substrates in an ultra-high vacuum (10$^{-8}$ Torr) magnetron sputtering system. The STO substrate was preliminarily 
annealed in vacuum for 1\,h at 600\,$^\circ$C, so as to remove the
surface contaminants. Subsequently, a Fe$_3$Ga$_4$ film 
was sputtered from the alloy target in a 3\,mTorr argon atmosphere at 600\,$^\circ$C.
Then the film was annealed \textit{in situ} for extra 2 hours to
improve its crystallinity. Finally, a Pt capping layer  
was sputtered on top of the
Fe$_3$Ga$_4$ layer in a 2\,mTorr argon atmosphere at room temperature.
The thickness of each layer was calibrated by means of x-ray reflectivity (XRR).\\

\noindent\textbf{Structural characterization}\\ 
The crystal structure and epitaxial nature of the Fe$_3$Ga$_4$
films were characterized by room-temperature x-ray diffraction measurements. 
The 2$\theta$-$\omega$ scan, XRR, and RSM were collected using a PANalytial X'Pert Pro x-ray diffractometer with 
Cu K$\alpha$ radiation ($\lambda$ = 1.5418\,\AA{}).  
The cross-sectional transmission electron microscopy (TEM) sample was
prepared using a focused ion beam and was measured by scanning electron microscope (Helios G4 UX, Thermo Fisher). Atomic resolution high-angle annular dark field (HAADF) scanning transmission electron microscopy (STEM) observations were performed using a 300\,kV spherical aberration (Cs)-corrected STEM (JEM-ARM300F, JEOL). The convergence angle was 28\,mrad and the angular range of %\tcr{the}
collected electrons for the HAADF-STEM imaging was 64--180\,mrad. The HAADF-STEM image was simulated using the Dr.\ Probe software package~\cite{Barthel2018}.\\

\noindent\textbf{Magnetization measurement}\\ 
The temperature- $M(T)$ and field-dependent $M(H)$ magnetizations
were measured using a Quantum Design magnetic property measurement system.
A Fe$_3$Ga$_4$ film with typical dimensions of 1.6 $\times$ 1.8\,mm$^2$
was used for the magnetization measurements.
The $M(T)$ data in various magnetic fields were collected using a
zero-field-cooling protocol. While for the $M(H)$ data, the full magnetic hysteresis loops were collected. For both measurements, the magnetic fields were applied both along the $b$ axis (i.e., perpendicular to the film plane) and within the $ac$ plane (i.e., parallel to the film plane). 
The signal from the STO substrate was subtracted from the measured $M(H)$ (see Supplementary Fig.~S4). While for the $M(T)$, considering that the STO substrate exhibits an almost temperature-independent magnetization, such subtraction was not
necessary and the data shown in fig.~S3 refer to the measured $M(T)$. \\

\noindent\textbf{Electrical transport measurement}\\	
For the transport measurements, the Fe$_3$Ga$_4$ film was patterned into a Hall-bar geometry (central area: 100\,{\textmu}m $\times$ 20\,{\textmu}m; electrodes: 20\,{\textmu}m $\times$ 20\,{\textmu}m) by photolithography and Ar-ion-beam etching techniques (see Fig.~\ref{fig:pH}a). The transport measurements were performed in a Quantum Design physical property measurement system using a standard four-probe method. 
For the resistivity measurements, the electric current was applied
in the film plane, while the magnetic field was applied perpendicular to the film plane. To avoid spurious resistivity contributions due to
misaligned Hall probes, all the resistivity measurements were performed in both positive- and negative magnetic fields. Then, in the case of the Hall resistivity $\rho_\mathrm{xy}$, the spurious longitudinal contribution was removed by an antisymmetrization procedure, i.e., $\rho_\mathrm{xy}(H) =[\rho_\mathrm{xy}(H)-\rho_\mathrm{xy}(-H)]/2$. Whereas in the case of the longitudinal electrical resistivity $\rho_\mathrm{xx}$, the spurious transverse contribution was removed by a symmetrization procedure, i.e., $\rho_\mathrm{xx}(H) =[\rho_\mathrm{xx}(H)+\rho_\mathrm{xx}(-H)]/2$. \\

\noindent\textbf{Analysis of Hall resistivity}\\ 
In magnetic materials with topologically chiral spin textures, the Hall resistivity is typically written in the form of 
$\rho_\mathrm{xy} = \rho_\mathrm{xy}^\mathrm{O} + \rho_\mathrm{xy}^\mathrm{A} +  \rho_\mathrm{xy}^\mathrm{U}$, where the three terms represent the contributions from the ordinary, the anomalous, and the THE, respectively. 
The ordinary Hall term $\rho_\mathrm{xy}^\mathrm{O}$ (= $R_\mathrm{0}$$H$) is proportional to the  applied magnetic field, the anomalous Hall term $\rho_\mathrm{xy}^\mathrm{A}$ (= $R_\mathrm{S}$$M$) is mostly determined by the magnetization. Here, $R_0$ is the ordinary Hall constant (that is closely related to the sign and density of the conducting carriers) and 
$R_\mathrm{S}$ is proportional to $\rho_\mathrm{xx}^2$ or $\rho_\mathrm{xx}$ or a constant. In the FM state, we simply use $M(H)$ to obtain  $\rho_\mathrm{xy}^\mathrm{A}$ from $\rho_\mathrm{xy}^\mathrm{A}=R_\mathrm{S}M$, where $R_\mathrm{S}$ is a field-independent scaling factor. In the AFM state, since the intrinsic mechanism is thought to
describe the Fe$_3$Ga$_4$ film in the
resistivity range, the relation $R_\mathrm{S}$ = $S_\mathrm{H}$$\rho_\mathrm{xx}^2(H)$ was used to interpret the $\rho_\mathrm{xy}^\mathrm{A}$ term, 
where $S_\mathrm{H}$ is the intrinsic AHE coefficient, a parameter with
a close connection to the electronic structure. We also note that the use of $M(H)$ or $\rho_\mathrm{xx}^2$$M(H)$ for $\rho_\mathrm{xy}^\mathrm{A}$ in the full temperature range leads to comparable $\rho_\mathrm{xy}^\mathrm{U}$ values (see Supplementary Fig.~S13). After subtracting both the ordinary $\rho_\mathrm{xy}^\mathrm{O}(H)$ and the anomalous term $\rho_\mathrm{xy}^\mathrm{A}(H)$ from the measured Hall resistivity $\rho_\mathrm{xy}(H)$, the unconventional anomalous Hall term $\rho_\mathrm{xy}^\mathrm{U}(H)$ was obtained. Further, to understand the nature of AHE in Fe$_3$Ga$_4$ films, the electrical- and Hall conductivity were calculated using the expressions $\sigma_\mathrm{xx} = \rho_\mathrm{xx}/(\rho_\mathrm{xx}^2 + \rho_\mathrm{xy}^2)$ and $\sigma_\mathrm{xy} = \rho_\mathrm{xy}/(\rho_\mathrm{xx}^2 + \rho_\mathrm{xy}^2)$, respectively. The 
anomalous Hall conductivity instead was calculated according to $\sigma_\mathrm{xy}^\mathrm{A} = \rho_\mathrm{xy}^\mathrm{A}/[(\rho_\mathrm{xy}^\mathrm{A})^2 + \rho_\mathrm{xx}^2]$, where $\rho_\mathrm{xx}$ is the zero-field electrical resistivity.\\

\noindent\textbf{Magnetic force microscopy}\\ 
Magnetic force microscopy experiments were carried out in a Bruker Icon atomic force microscope
using commercially available magnetic tips (Bruker, MESP-V2) with a typical spring constant of 2.8\,N/m. 
The MFM images were acquired 
using the lift mode, where the tip was lifted about 60\,nm above the sample surface.
In these MFM measurements, the phase shift of the cantilever is proportional to the out-of-plane stray-field gradient. %In the MFM images, the red (blue) regions represent up (down) magnetic domains, whose magnetization is parallel (antiparallel) to the magnetization direction of the tip (here coinciding with the applied magnetic field direction).
The NanoScope Analysis software allowed us to binarize all the images at the same threshold starting from the same vertical scale. The particle analysis module of this software was utilized to count the number of bubbles. The magnetic field was first set to zero before acquiring the MFM images at the target magnetic fields.

\vspace{10pt}
\noindent\textbf{Data availability}\\
All data needed to evaluate the conclusions in the paper are present in the paper and the Supplementary Information file. All raw data related to the current study are available from the corresponding authors upon request.

%\vspace{10pt}
%\noindent\textbf{Data and materials availability:} All the data needed to evaluate the reported conclusions 
%are presented in the paper and/or in the Supplementary Material. 
%Additional data related to this paper may be requested from the 
%authors. 

%\vspace{10pt}
%\noindent\textbf{Supplementary Materials}\\
%\textbf{This PDF file includes:}\\
%    Fig.~S1 to S13\\ %
%	Table S1\\

\vspace{10pt}
\noindent\textbf{References}

\bibliographystyle{naturemag}
\bibliography{FeGa}

\vspace{10pt}
\noindent\textbf{Acknowledgments}\\
We thank Xiaodong Zhou for fruitful discussions.
This work was supported by the National Natural Science Foundation of China (Grant Nos. 12374105 and 12350710785),
the Natural Science Foundation of Shanghai (Grant Nos.\ 21ZR1420500 and 21JC\-140\-2300), the Natural Science Foundation
of Chongqing (Grant No.\ CSTB-2022NSCQ-MSX1678), and the Fundamental Research Funds for the Central Universities. J.M.\ acknowledges the financial support from the National Natural Science Foundation of China (Grant Nos.\ U2032213 and 12334008).\\

\vspace{10pt}
\noindent\textbf{Author contributions}\\ 
T.Sha. and Q.Z. conceived and led the project. J.Me. and D.J.G. synthesized the sample. Z.Y. and Y.S. performed the scanning transmission electron microscopy measurements. K.Z. and H.W. fabricated the devices. H.Y. performed the magnetic force microscopy measurements. J.Me., Y.W., K.X., B.Y., X.Z., B.L., Y.H., J.Ma., and Y.X. performed the magnetization and electrical transport measurements. J.Me. and H.Y. analyzed the experimental data. J.Me., T.Shi., Q.Z., and T.Sha. wrote the paper with input from all authors.\\

\vspace{10pt}
\noindent\textbf{Conflict of interest}\\
The authors declare no competing interests.

\vspace{10pt}
\noindent\textbf{Additional information}\\
\noindent\textbf{Supplementary information} The online version contains supplementary material available at https://doi.org/xxxxx.

\end{document}